
\def\item{\par\indent\indent \hangindent3\parindent \textindent}

\newbox\partialpage
\def\begindoublecolumns{\begingroup
 \output={\global\setbox\partialpage=\vbox{\unvbox255\bigskip}}\eject
 \output={\doublecolumnout} \hsize=14pc \vsize=89pc}

\def\doublecolumnout{\splittopskip=\topskip\splitmaxdepth=\maxdepth
 \dimen1=144pc \advance\dimen1 by-\ht\partialpage
 \setbox0=\vsplit255 to\dimen1 \setbox2=\vsplit255 to\dimen1
 \shipout\pagesofar \unvbox255 \penalty\outputpenalty}
\def\pagesofar{\unvbox\partialpage
 \wd0=\hsize \wd2=\hsize \hbox to\pagewidth{\box0\hfil\box2}}
\def\balancecolumns{\setbox0=\vbox{\unvbox255} \dimen1=\ht0
 \advance\dimen1 by\topskip \advance\dimen1 by-\baselineskip
 \divide\dimen1 by2 \splittopskip=\topskip
 {\vbadness=10000 \loop \global\setbox3=\copy0
  \global\setbox1=\vsplit3 to\dimen1
  \ifdim\ht3>\dimen1 \global\advance\dimen1 by1pt \repeat}
 \setbox0=\vbox to\dimen1{\unvbox} \setbox2=\vbox to\dimen1{\unvbox3}
\pagesofar}

\def\cosmo{\input head.tex \twelvepoint \cosmo}

\def\boxit#1{\vbox{\hrule{\vrule\kern3pt
  \vbox{\kern3pt#1\kern3pt}kern3pt\vrule}\hrule}}

\newbox\bigstrutbox
\setbox\bigstrutbox=\hbox{\vrule height12pt depth5pt width0pt}
\def\bigstrut{\relax\ifmmode\copy\bigstrutbox\else\unhcopy\bigstrutbox\fi}

\def\la{\mathrel{\mathpalette\fun <}}
\def\ga{\mathrel{\mathpalette\fun >}}

\def\fun#1#2{\lower3.6pt\vbox{\baselineskip0pt\lineskip.9pt
  \ialign{$\mathsurround=0pt#1\hfil##\hfil$\crcr#2\crcr\sim\crcr}}}

\font\tenrm=cmr10 scaled 1200
\font\sevenrm=cmr7 scaled 1200
\font\fiverm=cmr5 scaled 1200
\font\teni=cmmi10 scaled 1200
\font\seveni=cmmi7 scaled 1200
\font\fivei=cmmi5 scaled 1200
\font\tensy=cmsy10 scaled 1200
\font\sevensy=cmsy7 scaled 1200
\font\fivesy=cmsy5 scaled 1200
\font\tenex=cmex10 scaled 1200
\font\tenbf=cmbx10 scaled 1200
\font\sevenbf=cmbx7 scaled 1200
\font\fivebf=cmbx5 scaled 1200
\font\tenit=cmti10 scaled 1200
 \font\tensl=cmsl10 scaled 1200
\font\tentt=cmtt10 scaled 1200

\skewchar\teni='177 \skewchar\seveni='177 \skewchar\fivei='177
\skewchar\tensy='60 \skewchar\sevensy='60 \skewchar\fivesy='60

\font\ninei=cmmi10
\font\sixi=cmmi7
\font\fouri=cmmi5
\font\ninesy=cmsy10
\font\sixsy=cmsy7
\font\foursy=cmsy5

\skewchar\ninei='177 \skewchar\sixi='177 \skewchar\fouri='177
\skewchar\ninesy='60 \skewchar\sixsy='60 \skewchar\foursy='60
  \def\twelvepoint{\def\rm{\fam0 \tenrm}
  \textfont0=\tenrm \scriptfont0=\sevenrm \scriptscriptfont0=\fiverm
  \rm
  \textfont1=\teni \scriptfont1=\seveni \scriptscriptfont1=\fivei
  \def\mit{\fam1 } \def\oldstyle{\fam1 \teni}
  \textfont2=\tensy \scriptfont2=\sevensy \scriptscriptfont2=\fivesy
  \def\cal{\fam2 }
  \textfont3=\tenex \scriptfont3=\tenex \scriptscriptfont3=\tenex
  \textfont\itfam=\tenit \def\it{\fam\itfam\tenit}
  \textfont\slfam=\tensl \def\sl{\fam\slfam\tensl}
  \textfont\bffam=\tenbf \scriptfont\bffam=\sevenbf
     \scriptscriptfont\bffam=\fivebf \def\bf{\fam\bffam\tenbf}
  \textfont\ttfam=\tentt \def\tt{\fam\ttfam\tentt}
  \baselineskip=14pt}

\outer\def\beginsection#1\par{\bigskip\vbox{\message{#1}\noindent{\bf#1}}
  \nobreak\smallskip\vskip-\parskip\noindent}
\def\exdent#1\par{\noindent\hang\frenchspacing#1\par}
\def\ref {\par\noindent \hangindent=20pt \hangafter=1}
\hsize=6.5in
\vsize=9in
\parskip=0pt

\def\m@th{\mathsurround=0pt}
\def\complex#1{\left\{\,\vcenter{\baselineskip=18pt\m@th
   \ialign{$\displaystyle{##}\hfil$&\quad$\displaystyle{##}\hfil$\crcr#1
   \crcr}}\right.}


\def\leaderfill{\leaders\hbox to 1em{\hss.\hss}\hfill}
\twelvepoint
\baselineskip=18pt
\space\
\vskip 0.8in
\centerline{\bf A TYPE II SUPERNOVAE CONSTRAINT ON $\nu_e$-$\nu_s$
MIXING}
\vskip 1.2cm
\centerline{{\bf X. Shi} and {\bf G. Sigl}}
\vskip 0.25cm
\centerline{\sl Astronomy \& Astrophysics Center,}
\centerline{\sl The University of Chicago,}
\centerline{\sl Chicago, IL 60637-1433}
\vskip .4cm
\vskip 0.5in
\centerline{ABSTRACT}
\vskip 0.23in
\noindent The role of a resonant $\nu_e$-$\nu_s$ oscillation is discussed
in the event of a supernova explosion. It is concluded that
a significant $\nu_e$-$\nu_s$ mixing may hinder the ability of the supernova
to explode. It may also cool the proto-neutron
star too quickly with respect to the observed cooling time of several
seconds. The constraints on the $\nu_e$-$\nu_s$
mixing parameters based on the above arguments are calculated.
\vskip 2.9in


\vfill\eject
\centerline{\bf 1. Introduction}
\vskip 0.25cm
Oscillations between active neutrinos $\nu_{\alpha}$
and sterile neutrinos $\nu_s$ have
been discussed in various astrophysical and cosmological environments.
Such examples include the oscillation between $\nu_e$ and $\nu_s$ as
solutions to the solar neutrino problem [1], and oscillations between
$\nu_{\alpha}$ and $\nu_s$ in Big Bang Nucleosynthesis which may affect
the primordial $^4$He abundance [2-6].
Constraints on neutrino mixing parameters
have been obtained by requiring the predicted $^4$He abundance under these
neutrino mixings not to exceed the observed value [3-5].
Sterile neutrinos are also suggested as dark matter candidates if they have
proper mixings with active neutrinos [6].
In this paper we attempt to consider $\nu_e$-$\nu_s$ mixing in another
astrophysical environment--the Type II supernova (hereafter we
omit ``type II'' for simplicity) core.

Since there isn't a fully convincing numerical
supernova model at present, we try to
construct our arguments as model-insensitively as possible.
Supernova explosions occur when massive stars ($> 8 M_\odot$) reach
the end of their evolution [7]. For such massive stars,
an iron core is developed at the center of the
progenitor star, surrounded sequentially by shells of silicon, oxygen,
nitrogen,
carbon, helium, and finally an envelope of hydrogen. Nuclear reactions
occur at the interfaces of these shells and between the shell and the core.
The initial
iron core of the progenitor star has a density of about $10^9$ g/cm$^3$,
and is primarily supported by the pressure of the degenerate electron gas.
As the nuclear burning of the silicon shell continues,
more iron is deposited into the core,
until it reaches the Chandrasekhar mass ($\approx 1.4 M_\odot$),
when the core cannot be sufficiently supported by the electron
degeneracy pressure and begins to collapse. In the meantime, due to the high
density and the high temperature of the core,
the deleptonization process $e^-+p\rightarrow n+\nu_e$
occurs rapidly. When the density reaches
$10^{11}$ g/cm$^3$, the mean free path of neutrinos $l$ ($\approx
10^3/\rho_{14}$ cm for neutrinos of $\sim$ 20 MeV, where $\rho_{14}$
is the core density divided by $10^{14}$ g/cm$^3$)
is shorter than the radius of the core $R_{\rm core}\sim 10-100$km.
When the density is higher than
$10^{12}$ g/cm$^3$, the diffusion timescale of neutrinos becomes
longer than the timescale of the collapse of the core.
Neutrinos are thereafter trapped inside the core instead of escaping freely.
The collapse of the core finally stalls and bounces back
when the center of the core reaches approximately nuclear density
few$\times 10^{14}$
g/cm$^3$, at which the pressure of the degenerate neutron gas stops the
gravitational collapse. The bounce creates a shock wave in the midst
of the core that propagates outwards through the infalling matter.
It is generally believed that if the core is small ($< 1.4 M_\odot$),
the shock wave will have enough energy to blow
the matter outside the core away and lead to an explosion event [8].
If the core is sufficiently massive,
however, the shock wave quickly stalls within the core by losing energy
in dissociating large quantities of infalling nuclei into free nucleons.
It is currently believed that in this case
neutrinos deposit their energies into the shock several tenths of a second
after the bounce and reheat the
shock wave to eventually blow the outer layer away [9].

It is estimated that about 99$\%$ of the energy is emitted in the form
of neutrino bursts. Therefore, the role of neutrinos in the supernova
event is essential [9]. At the end of the collapse,
electron neutrinos are trapped instead of streaming out of the core.
The $\nu_\mu$, $\nu_\tau$, and anti-neutrinos
are also produced thermally inside the hot core with
an average energy of several tens of MeV.
This leads to a neutrino sphere with an average energy of $\sim 20$ MeV
trapped inside a dense core of nucleons and heavy nuclei. The number density
of $\bar\nu_e$ is highly suppressed at the collapsing
stage of the supernova due to the existence of the
degenerate $\nu_e$ gas. Its number density builds up only after
significant amount of $\nu_e$ escapes from the core after deleptonization.
It was shown by Wilson {\sl et al.} that the $\nu$ and $\bar\nu$
annihilating at the neutrino sphere can deposit their energies
in the shock and lift the shock front outwards, thus leading to an explosion
[10]. Diffusive neutrinos further cool down the core (a proto-neutron star)
to a neutron star after the explosion,
on a timescale of about 10 seconds. This diffusive $\bar\nu_e$ flux has
been observed by Kamioka [11] and IMB [12] in conjunction with SN1987A,
with a typical energy of 15--20 MeV and a timescale
of about 10 seconds. Therefore, the current picture of a
supernova event has received significant observational support.
\vskip 0.25cm
\centerline{\bf 2. Neutrino Mixing in the Supernova Core}
\vskip 0.25cm
The above picture assumes the standard electroweak theory.
If neutrino mixing is present, the picture may change substantially.
Many literatures have been focusing on the constraints
on properties of neutrinos [13].
We will focus on the impact of a $\nu_e$-$\nu_s$ mixing on the
supernova event, which hasn't been explored in full detail before [14].
In particular, instead of discussing the general cases of
the mixing, we will discuss only the case of
a resonant oscillation between $\nu_e$ and $\nu_s$,
which are relatively less model-dependent.

In the flavor basis the effective Hamiltonian of the
$\nu_\alpha$ (active)-$\nu_s$ (sterile) mixing system in a
medium is given by [15]
$$H={1\over 4E}
\left(\matrix{-\delta m^2\cos2\theta+4EV_{\nu_\alpha}&\delta m^2\sin2\theta\cr
 \delta m^2\sin2\theta&\delta m^2\cos2\theta\cr}\right),\eqno (1)$$
where $E$ is the neutrino energy and $\delta m^2=m_2^2-m_1^2$ is
the difference of the squares of the two vacuum mass eigenvalues
$m_1$  and $m_2$. We assume $\delta m^2>0$ when the sterile
neutrino is heavier. Furthermore, $\theta$ is the vaccum mixing
angle and $V_{\nu_\alpha}$ is the effective potential of the active
neutrino $\nu_\alpha$ in the medium. In a stellar environment [16],
\vfill\eject
$$\eqalignno{
V_{\nu_e}&=\sqrt{2}G_{\rm F}(N_e-0.5N_n+2N_{\nu_e}+
N_{\nu_\mu}+N_{\nu_\tau}) ,\cr
V_{\nu_\mu}&=\sqrt{2}G_{\rm F}(-0.5N_n+N_{\nu_e}+
2N_{\nu_\mu}+N_{\nu_\tau})& (2)\cr}$$
where $N_e$, $N_n$ and $N_{\nu_\alpha}$
are the number densities of electrons,
neutrons, and neutrinos minus the number densities of their
anti-particles, respectively.
$V_{\nu_\tau}$ is similar to $V_{\nu_\mu}$ except that $\nu_\mu$ and
$\nu_\tau$ are interchanged in the equation. The effective
potential of a sterile neutrino is always zero.
Based on $CP$ symmetry, $V_{{\bar\nu}_\alpha}=-V_{\nu_\alpha}$.
In the special situation of a supernova core, $N_{\nu_\mu}$ and $N_{\nu_\tau}$
are negligible, since $\nu_\mu$ and $\nu_\tau$ are only produced in
$\nu\bar\nu$ pairs via neutral current processes.

A resonance between the two mixing neutrinos occurs
when the diagonal elements of the effective Hamiltonian eq.~(1)
become equal, i.e., when the neutrino energy
equals the resonance energy $E_r$ given by
$$E_r={\delta m^2\cos 2\theta\over 2V_{\nu_\alpha}}.\eqno(3)$$
Inside a supernova core, where spherical symmetry is a reliable approximation,
the region where such a resonance occurs for neutrinos with an
energy $E_r$ is a spherical shell with radius $R_{\rm res}(E_r)$.
It can be shown that if the sterile neutrinos resonantly converted from
the active neutrinos escape freely from the core (we will discuss
the condition for this later),
the usual continuity equation relating the
local rate of change of the active neutrino number density
$n_{\nu_\alpha}$ to the divergence of the corresponding local
flux ${\vec\phi}_{\nu_\alpha}$ is modified to
$${\dot n}_{\nu_\alpha}=-\nabla\cdot{\vec\phi}_{\nu_\alpha}
  -\left[\nabla E_{\rm r}\cdot{d{\vec\phi}^{\rm out}_{\nu_\alpha}
  \over dE}({E_r})
  +\dot{E_{\rm r}}{d n^{\rm out}_{\nu_\alpha}\over dE}({E_r})\right]
  P_{\nu_\alpha}.\eqno (4)$$
Here, ${\vec\phi}^{\rm out}$ and
$n^{\rm out}_{\nu_\alpha}$ are the partial flux and
number density which only account for $\nu_{\alpha}$'s going outward
with respect to the surface of constant resonance energy $E_r$ (outward
radial direction in our case).
Furthermore, $P_\alpha$ is the probability for an active
neutrino going through the resonance to be converted to the
sterile neutrino $\nu_s$. We have neglected terms
corresponding to $\nu_s\to\nu_\alpha$ convertions and
$\nu_\alpha\to\nu_s$ convertions of inward going $\nu_\alpha$,
since we are only interested in the
global $\nu_\alpha$ loss by integrating
eq.~(4) over the whole SN core to which the sum of the neglected terms does
not contribute.

In general, $P_\alpha$ depends on the
rate of decoherence of the neutrino flavor density matrix caused
by non-forward scattering off the background [14] as well as on
the adiabaticity of the resonance. In the limiting case where
the width of the resonance region is much shorter
than the mean free path of neutrinos, which is the case we will do our
calculations,
$$P_\alpha(E_r)\approx 1-\exp\Bigl[{-\pi\delta m^2\sin^22\theta
\over 4\cos2\theta}\Bigl\vert\nabla E_r+{\dot{E_r}\over c}\Bigr\vert
^{-1}\Bigr].\eqno (5)$$
This is correct to the first order in the change of $E_r$ in the
resonance region [15]. $P_\alpha\sim 1$ corresponds to a
highly adiabatic resonance whereas $P_\alpha\ll 1$
represents a highly non-adiabatic resonance.
On the other hand, when the width of the resonance region
is much larger than the neutrino
mean free path, $P_\alpha\sim 0.5$ [14], which means the resonance is always
``adiabatic''. Therefore, it can be seen from our following calculations that
our constraints that exclude sufficiently adiabatic
resonant mixings will equally apply to the case with large
resonance width, as long as the $\nu_s$ stream freely outside of
the resonance region.

To get an order of magnitude estimate of the neutrino loss rate,
we can further simplify
eq.~(4) by setting $\nabla E_r\cdot(d{\vec\phi}^{\rm out}/
dE)(E_r)\approx (dE_r/dr) ncf(E_r)E^2_r/\overline E^3$ and
$(dn^{\rm out}/dE)(E_r)\approx nf(E_r)E^2_r/\overline E^3$
(we drop the subscript $\nu_{\alpha}$ from now on). Here,
$c$ is the speed of light, $f(E_r)$ is the occupation number at a state
with $E_r$, and $\overline E$ is the average neutrino
energy which for the partially degenerate $\nu_e$
is of the order of its chemical potential $\mu$ and for other neutrino species
(and all antineutrinos) is 3 times their temperatures in the supernova core.
We can then integrate
the second term on the r.h.s. of eq.~(4) over the SN core
to get the additional global $\nu_{\alpha}$ loss rate $\dot N_{\rm res}$
and the additional global energy
loss rate $\dot Q_{\rm res}$ due to the resonant conversion,
$$\dot N_{\rm res}\approx -\int_0^{R_{\rm core}}4\pi r^2 n{dE_r\over
dr}\left(c+ v_r\right)f(E_r){E_r^2\over{\overline E}^3}P(E_r)dr,$$
$$\dot Q_{\rm res}\approx -\int_0^{R_{\rm core}}4\pi r^2 n{dE_r\over
dr}\left(c+ v_r\right)f(E_r)\left({E_r\over\overline E}\right)^3
P(E_r)dr,\eqno (6)$$
where $v_r=\dot E_r/\vert\nabla E_r\vert$, the speed
of the resonance shell for neutrinos with energy $E_r$.
The integrals on the r.h.s. are dominated by integrands at $E_r\sim\overline E$
if the resonance for neutrinos with energy $\overline E$ ever
occurs inside the core.
When $E_r\sim\overline E$, we always expect $\vert\dot E\vert<<c\cdot dE/dr$
since otherwise the neutrinos will build up inside or dissipate from
the core in $R_{\rm core}/c\sim 10^{-4}$s, which contradicts our
knowledge of supernovae. Therefore $v_r$ and $\dot E$ are negleted in the
following discussion.
If we denote $N_{\rm tot}$ and $Q_{\rm tot}$
as the total number and the total energy of $\nu_\alpha$ inside the core,
we can approximate eq.~(6) by
$$\dot N_{\rm res}\sim -{cP(\overline E)\vert_{\dot E_r=0}
\over R_{\rm core}}N_{\rm tot},
\quad\dot Q_{\rm res}\sim -{cP(\overline E)\vert_{\dot E_r=0}
\over R_{\rm core}}Q_{\rm tot}.\eqno (7)$$

One crucial quantity in the problem is the scale height of $E_r$,
$\vert d\ln E_r/dr\vert ^{-1}$. A rough estimate for it is
$$\Bigl\vert{d\ln E_r\over dr}\Bigr\vert^{-1}
=\Bigl\vert{d\ln N_{\rm eff}\over dr}\Bigr\vert^{-1}
\sim\Bigl\vert{d\ln n\over dr}\Bigr\vert^{-1}\Bigl\vert{N_{\rm eff}\over n}
\Bigr\vert,\eqno(8)$$
where $\vert{d\ln n/dr}\vert^{-1}$ is the scale height of the neutrino
density $n$ at the resonance region, and
$N_{\rm eff}=V_{\nu_\alpha}/\sqrt{2}G_{\rm F}$.
Without going into detailed models,
$$\Bigl\vert{d\ln n\over dr}\Bigr\vert^{-1}>l.\eqno (9)$$

In the following sections we apply these results to estimate the
loss rates in the collapsing and the cooling phases of supernovae.
\vskip 0.25cm
\centerline{\bf 3. Constraints from the Collapsing Phase}
\vskip 0.25cm
During the collapsing phase of supernovae, $\nu_e$ will encounter
a resonance when the $\nu_e$-$\nu_s$ mixing satisfies
$$\delta m^2\cos2\theta=2\sqrt{2}G_{\rm F}E_rN_{\rm eff}
=2\sqrt{2}G_{\rm F}E_r(N_e-0.5N_n+2N_{\nu_e}).\eqno(10)$$
Taking $\overline E\sim 10$MeV, and
the density of the core
$\rho\sim 10^{12}$ g/cm$^3$ to several times $10^{14}$ g/cm$^3$,
we find 10$^6$ eV$^2\la\delta m^2\la 10^8$ eV$^2$.
At this stage the $\nu_e$'s are already trapped in the SN core.
The relevant time scale here is the
free fall time scale of about $10^{-3}$s$\sim R_{\rm core}/(0.1c)$.
Then if $\dot N_{res}>(0.1c/R_{\rm core})N_{\rm tot}$, $\nu_e$ will never
get trapped after the central density of the core
reaches and exceeds the resonant value. The process is illustrated in
Figure 1. It has been shown that no explosion occurs when the trapped
lepton number is less than 0.9 times the lepton number before the
collapse [17]. Therefore we require
$$P(\overline E)\vert_{\dot E=0}\la 0.1\quad{\rm for}\quad
10^6{\rm eV}^2\la\delta m^2\la 10^8{\rm eV}^2\eqno(11)$$
so that no serious leakage of $\nu_e$ will occur.
Because of eq.~(5) this corresponds to
$${\delta m^2\sin^22\theta\over\overline E}\Bigl\vert {d\ln
E_r\over dr}(\overline E)\Bigr\vert^{-1}\la 0.1.\eqno (12)$$
Since $N_{\rm eff}\sim n$,
$\vert{d\ln E_r/dr}(\overline E)\vert^{-1}>l\sim 10^3$cm from
eqs.~(8) and (9). Taking $\overline E\sim 10$MeV,
we conclude that $\nu_e$-$\nu_s$ mixings with
$$\delta m^2\sin^22\theta\ga 10^{-2} {\rm eV}^2\quad {\rm and}\quad
10^6{\rm eV}^2\la \delta m^2\la 10^8{\rm eV}^2\eqno (13)$$
are ruled out so that the supernova core can trap
enough lepton number to explode.
\vskip 0.25cm
\centerline{\bf 4. Constraints from the Cooling Phase}
\vskip 0.25cm
For $\delta m^2\la 10^6$ eV$^2$, interesting constraints on the
$\nu_e$-$\nu_s$ mixing come from the consideration of the
resonance that occurs after the bounce.
After the bounce, $N_{\rm eff}=
N_e-0.5N_n+2N_{\nu_e}$ drops below zero very quickly at the outer part of
the core [17,18].
(In general $N_e=N_p\approx N_n$ before the collapse, where
$N_p$ is the number density of protons. After about 1/3 of protons
are deleptonized into neutrons, $N_{\rm eff}$ will be negative.)
However, at the center of the core, there is a significant
build-up of $\nu_e$, $N_{\rm eff}$ remains positive.
Figure 2 shows roughly the relation between $N_{\rm eff}$ and the
position inside the core at about 0.5 seconds after the bounce, as
inferred from figure 10f of ref. 18 (similar relations exist for
prompt-explosion models [17]).
Thus, there exists a central part of the core with a positive $V_{\nu_e}$
and an outer part of the core with a negative $V_{\nu_e}$. Furthermore,
outside the core where no significant deleptonization
has taken place, $V_{\nu_e}$ is still positive.
Therefore two resonance regions may form, (for both $\nu_e$-$\nu_s$ mixings
and $\bar\nu_e$-$\bar\nu_s$ mixings), one inside the core and one
outside the core, with
$$N_{\rm eff}\approx 0,\eqno (14)$$
if $\delta m^2$ is much smaller than $G_{\rm F}EN_{\rm eff}$ and
can be neglected.
If we take $\overline E\sim 10$ MeV at the edge of the neutrino sphere,
we can see from figure 2 that when $\delta m^2\la 10^6$ eV$^2$, both $\nu_e$
and $\bar\nu_e$ can be resonantly converted into sterile neutrinos inside
the core.

In this section we consider the resonance region inside the core. In this case,
the scale height of $E_r$ is much smaller than in the previous section due
to eqs.~(8) and (14). By using eq.~(10) and $n\la\overline E^3/6\pi^2$,
we find for $E_r\sim \overline E$,
$$\Bigl\vert{d\ln E_r\over dr}\Bigr\vert^{-1} (\overline E)
>l\Bigl\vert{N_{\rm eff}\over n}\Bigr\vert
\ga l{3\pi^2\delta m^2\over\sqrt 2G_{\rm F}\overline E^4}.\eqno (15)$$

The concern is that if the resonance occurs with a sufficient
adiabaticity, the intensive production of the freely streaming
sterile neutrinos will cool down the core
too quickly compared with the observed cooling time of about 10
seconds. This translates into
$${cP\vert_{\dot E_r=0}\over R_{\rm core}}(\overline E)\gg 1\ {\rm sec}^{-1},
\quad{\rm or\ equivalently}\quad
P\vert_{\dot E_r=0}(\overline E)\ga 10^{-3}\eqno (16)$$
for $R_{\rm core}\sim 10$ to 100km. By using an approximation
$$P\vert_{\dot E_r=0}(\overline E)\sim
{\delta m^2\sin^22\theta\over\cos2\theta}\vert\nabla E_r\vert^{-1}\quad
{\rm for\ } P\ll 1, \eqno (17)$$
we conclude after combining eqs.~(15), (16) and (17) that if
$$\Bigl({\delta m^2\sin^22\theta\over\cos2\theta}\Bigr)
\Bigl({3\pi^2\delta m^2 l\over\sqrt 2G_{\rm F}\overline E^5}\Bigr)
\ga 10^{-3},\eqno (18)$$
the proto-neutron star will cool down much faster than the observed timescale
by losing neutrinos. Taking $\overline E\sim 50$ MeV, $l\sim 10^3$cm,
eq. (18) excludes $\nu_e$-$\nu_s$ mixings with
$$\delta m^2\sin2\theta\ga30{\rm eV}^2\quad{\rm and}\quad\delta
m^2\la 10^6{\rm eV}^2.\eqno(19)$$
\vskip 0.25cm
\centerline{\bf 5. Possible Tightening of the Bound by
Reconversion}
\vskip 0.25cm
Since the effective potential $V_{\nu_e}$ goes from negative at
the edge of the core to positive outside the core, at least the
$\bar\nu_e$-$\bar\nu_s$ will encounter a second resonance if
$\delta m^2\la 10^6$ eV$^2$.
The density outside the core is quite
low compared with that of the core. As a result, this outer
resonance could be highly adiabatic since $\vert d\ln E_r/dr\vert^{-1}$
could be much larger.
Then all the $\bar\nu_s$ converted from $\bar\nu_e$ deep in the core
will be completely reconverted into $\bar\nu_e$ which
then freely escape from the supernova because the outer resonance
is located above the neutrino sphere.
But these neutrinos have energies $\ga$50 MeV as they are
originated from $\bar\nu_e$'s deep inside the neutrino sphere.
Since the cross section of neutrinos in the $\nu$-e scattering
experiments (Kamiokande and IMB) scales roughly with $E^2$, where
$E$ is the energy of the neutrinos, the fact that we didn't see
any $\bar\nu_e$ with $E\ga 50$ MeV
in these experiments will put a tighter bound
on $\nu_e$-$\nu_s$ mixings than that of the previous section.
The average energy of $\bar\nu_e$ observed by Kamiokande and IMB
is $\sim 20$ MeV [13]. Therefore, the above consideration
could extend eq.~(16) by two order of magnitude, because even
if the energy loss rate due to the resonant conversion
is only one tenth of the standard loss rate,
$\sim 5$ of the 20 neutrinos seen by IMB
and Kamiokande should have had an energy of $\ga$50 MeV. More
generally, if in the future a neutrino detector detects $\cal N$
supernova neutrinos with none of them having an energy
$\ga$ 50 MeV, the excluded parameter region would extend to
$$\delta m^2\sin 2\theta\ga\left({30\over \cal N}\right)
{\rm eV}^2\quad{\rm and}\quad\delta m^2\la 10^6
{\rm eV}^2.\eqno(20)$$
With the current data, ${\cal N}=20$, the excluded region is
$$\delta m^2\sin 2\theta\ga 1{\rm eV}^2\quad{\rm and}\quad\delta m^2\la 10^6
{\rm eV}^2.\eqno(21)$$
One has to keep in mind, however, that eqs. (20) and (21) are only true if
the outer resonance is highly adiabatic, which for eq. (21) requires
the density scale height at the outer resonance to be larger than
$\sim (10/\sin2\theta)$m, which seems fairly reasonable when
compared with the core dimensions.

\vskip 0.25cm
\centerline{\bf 6. Discussion and Summary}
\vskip 0.25cm
As we mentioned in the previous section, the above calculations
are valid only if the sterile neutrinos have a mean free path larger than the
radius of the core after passing the resonance. Outside the resonance region,
the mean free path of sterile neutrinos is roughly
$l/\sin^2\theta_m$ [3--5,14], where $\theta_m$ is a typical
medium mixing angle. In the case of the constraints from the
collapsing phase discussed in section 3 this is of the same
order as the vacuum mixing angle $\theta$ whereas in the case of
section 4 due to medium effects this is even suppressed compared
to $\theta$. Our calculations are thus valid as long as
$$\sin^2\theta< l/R_{\rm core}\sim 10^{-3}.\eqno (22)$$

Figure 3 shows the excluded region on the $\nu_e$-$\nu_s$ mixing parameter
space from our consideration of a resonant $\nu_e$-$\nu_s$ mixing
in both the pre-bounce and the post-bounce supernova core, eqs.~(13) and
(19), as well as the possible tighter bound eq.~(21), and
eq.~(22) under which the above bounds are valid.
It also shows the parameter space excluded from Big Bang Nucleosynthesis,
which gives a stricter
constraints for $\delta m^2\la 10^3$ eV$^2$ [3--5].
Tha hatched region in Fig.~3 shows the required mixing
parameters for the sterile neutrino to contribute between 30\%
and 100\% of the critical density of the
universe [5,6] (assuming a Hubble constant between 50 km/sec/Mpc
and 100 km/sec/Mpc).
A significant part of the hatched region (at which the sterile
neutrino serves as warm dark matter)
is excluded by our constraints from the supernova.

In summary, we have considered the effect of a resonant $\nu_e$-$\nu_s$ mixing
on a type II supernova. By requiring that a supernova
retains enough leptons to explode and is not cooled down
within a fraction of a second, we obtain constraints on the $\nu_e$-$\nu_s$
mixing parameter space. The excluded region includes the region required
for the sterile neutrino to be the warm dark matter through
$\nu_e$-$\nu_s$ oscillations in the early universe.
\centerline{\bf Acknowledgement}
We thank D. N. Schramm for highly valuable suggestions.
This work was supported by DoE (Nuclear) and
NSF at the Unversity of Chicago, and by the NASA/Fermilab
astrophysics program as well as the Alexander-von-Humboldt foundation.
\vfill\eject
\noindent {\bf References:}

\noindent 1. D. N. Schramm and X. Shi, in {\sl Proc. TAUP'93},
Gran Sasso Laboratory, Italy, 1993, ed. A. Bottino, Elsevier Science
Publishers B.V. (to be published), and references therein.

\noindent 2. A. Dolgov, {\sl Sov. J. Nucl. Phys.}, {\bf 33}, 700 (1981);
R. Barbieri and A. Dolgov, {\sl Nucl. Phys.}, {\bf B 349}, 743 (1991).

\noindent 3. Enqvist {\sl et al.}, {\sl Nucl. Phys.} {\bf B373}, 498 (1992).

\noindent 4.  J. M. Cline, {\sl Phys. Rev. Lett.}, {\bf 68}, 3137 (1992).

\noindent 5. X. Shi, D. N. Schramm and B. D. Fields, {\sl Phys. Rev} D.,
{\bf 48}, 2563 (1993).

\noindent 6. Scott Dodelson and L. M. Widrow, FERMILAB-Pub-93/057-A.

\noindent 7. For a review on supernovae, see {\sl Supernovae}, ed.
A. G. Petschek, Springer-Verlag, 1990; or {\sl Supernovae, 10th
Santa Cruz Summer Workshop in Astronomy and Astrophysics},  ed.
S. E. Woosley, Springer-Verlag, 1989.

\noindent 8. M. Takahara and K. Sato, {\sl Prog. Theor. Phys.}, {\bf 71},
524 and {\bf 72}, 978 (1984).

\noindent 9. Section VII of {\sl Supernovae, 10th
Santa Cruz Summer Workshop in Astronomy and Astrophysics}, ed.
S. E. Woosley, Springer-Verlag, 1989.

\noindent 10. Ronald W. Mayle and James R. Wilson, in {\sl Supernovae, 10th
Santa Cruz Summer Workshop in Astronomy and Astrophysics}, ed.
S. E. Woosley, Springer-Verlag, 1989.

\noindent 11. K. Hirata {\sl et al.}, {\sl Phys. Rev. Lett.} {\bf 58},
1490 (1987).

\noindent 12. R. Bionta {\sl et al.}, {\sl Phys. Rev. Lett.} {\bf 58},
1494 (1987).

\noindent 13. D. N. Schramm and J. W. Truran, {\sl Physics Report},
{\bf 189}, No. 2, 89 (1990) and references therein.

\noindent 14. G. Raffelt and G. Sigl, {\sl Astroparticle Physics} {\bf 1},
165 (1993).

\noindent 15. T. K. Kuo and James Panteleone, {\sl
Rev. Mod. Phys.}, {\bf Vol. 61}, No. 4, 937 (1989).

\noindent 16. D. N\"otzold and G. Raffelt, {\sl Nuclear Physics} {\bf B 307},
924 (1988).

\noindent 17. S. W. Bruenn, {\sl Ap. J.} {\bf 340}, 955 (1989).

\noindent 18. James R. Wilson and Ronald W. Mayle, in {\sl The Nuclear Equation
of State, Part A}, eds. W. Greiner and H. St\=ocker, Plenum Press,
New York, 1989.

\vfill\eject
\noindent {\bf Figure Caption:}

\noindent Figure 1. An illustration of the configuration of
$V_{\nu_e}$ inside the core during the core collapse with (the dashed line)
and without (the solid line) the
$\nu_e$-$\nu_s$ mixing that satisfies eq.~(13).
\vskip 0.25cm
\noindent Figure 2. An illustration of $N_{\rm eff}$
inside the core at the bounce and several tenths of a second
after the bounce. $R_{ns}$ is the radius of
the neutrino sphere.
\vskip 0.25cm
\noindent Figure 3. On the $\nu_e$-$\nu_s$ mixing paramter space,
the area enclosed by the solid line
is excluded by our supernova consideration. The area enclosed by the
long-dashed line is the possible extended bound due to the resonance outside
the core. The area
enclosed by short-dashed line is excluded by Big Bang
Nucleosynthesis. The hatched region shows the parameters for
a sterile neutrino dark matter candidate (generated
a $\nu_e$-$\nu_s$ mixing in the early universe) to contribute
30\% to 100\% the critical density of the universe today (the Hubble
constant is taken to be between 50--100km/sec/Mpc).
\end